\documentstyle[epsfig]{aipproc}

\def\ltap{\raisebox{-.4ex}{\rlap{$\sim$}} \raisebox{.4ex}{$<$}}
\def\rts{\sqrt s}
\def\wp{W^+}
\def\wm{W^-}
\def\anti{\overline}
\def\fbi{~{\rm fb}^{-1}}

\def\mev{~{\rm MeV}}
\def\gev{~{\rm GeV}}

\def\anti{\overline}

\def\eg{{\it e.g.}}
\def\etal{{\it et al.}}

\def\epem{{e^+e^-}}

\def\br{B}

\def\h{h}
\def\mh{m_{\h}}
\def\mm{\mu^+\mu^-}
\def\mt{m_t}

\def\mw{M_W}
\def\mz{M_Z}

\def\lsim{\ltap}

\begin{document}
\hspace*{3.6in}{\bf IUHET-378}\\
\hspace*{3.6in}{\bf December 1997}
\title{Precision $W$-Boson and Higgs Boson\\ 
Mass Determinations at Muon Colliders\thanks{Presented
at the Workshop on Physics at the First Muon Collider and at the
Front End of a Muon Collider, November 6-9, 1997,
Fermi National Accelerator Laboratory.}}

\author{M. S. Berger}
\address{Indiana University\\
Bloomington Indiana 47405}

\maketitle

\begin{abstract}
Precise determinations of the masses  of the $W$ boson and of the top quark
could stringently test the radiative structure of the Standard Model (SM) or
provide evidence for new physics. We analyze the excellent prospects at
a muon collider for measuring $M_W$ and $m_t$
in the $W^+W^-$ and $ZH$ threshold regions. With an integrated
luminosity of 10 (100)~fb$^{-1}$,
the $W$-boson mass could be measured to
a precision of 20 (6)~MeV, provided that
theoretical and  experimental systematics are understood.
A measurement of 
$\Delta M_W=6$~MeV 
would constrain the mass of a  $\sim 100$~GeV Higgs  to about $\pm 10$~GeV.
We demonstrate that a measurement at future colliders of the 
Bjorken process.
With
an integrated luminosity of $100\fbi$ it is possible to measure
the Standard Model Higgs mass to within 45\mev\
at a $\mu^+\mu^-$ collider for $\mh=100\gev$.
\end{abstract}

\section*{Introduction}
%
%
%
%

Muon colliders offer a wide range of opportunities for exploring physics
within and beyond the Standard Model
(SM).
An important potential application of these machines is the precision
measurement of  particle masses, widths and couplings.
We estimate here the accuracy with which the $W$ and Higgs
boson masses
can be determined from $\wp\wm$ and $ZH$ threshold measurements
at a muon collider\cite{bbgh3,Zh}.

\section*{$M_W$ Measurement at the $\mu^+\mu^-\to W^+W^-$ Threshold}

A muon collider is particularly
well suited to the threshold measurement because the energy of the beam has
a very narrow spread. 
The threshold cross section\cite{mnw,coulomb} is most
sensitive to $\mw$ just above $\sqrt s = 2\mw$,
but a tradeoff exists between
maximizing the signal rate and the sensitivity of the cross section to $\mw$.
Detailed analysis \cite{lepii} shows that
if the background level is small and systematic uncertainties in efficiencies
are not important, then the optimal measurement of $\mw$
is obtained by collecting data at a single energy
$$\sqrt{s} \sim 2\mw + 0.5  \gev\ \sim\ 161  \gev ,$$
where the threshold cross section is sharply rising.

\begin{figure}[htb]
\leavevmode
\begin{center}
\epsfxsize=3.0in\hspace{0in}\epsffile{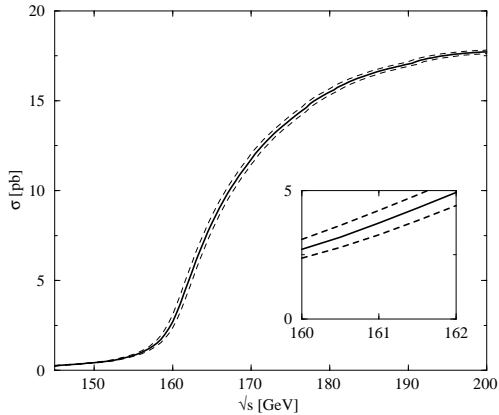}
\end{center}
\caption[]{\footnotesize\sf The cross section for $\mm \to W^+W^-$
in the threshold region for $\mw=80.3$~GeV (solid) and $\mw=80.1,80.5$~GeV
(dashed).
The inlaid graph shows the region of the threshold curve where the
statistical sensitivity to $\mw$ is maximized.
Effects of ISR have been included.}
\label{figure2}
\end{figure}

For a LEP2 measurement with 100~pb$^{-1}$ of integrated luminosity
the background and systematic uncertainties are, in fact, sufficiently small
that the error for $\mw$ will be limited by the statistical uncertainty
of the measurement at $\rts=161\gev$.  But, at a muon
collider at high luminosity, systematic errors
arising from uncertainties in the background level and the detection/triggering
efficiencies will be dominant unless
some of the luminosity is devoted to measuring
the level of the background (which automatically includes somewhat similar
efficiencies) at an energy below the $W^+W^-$ threshold.
Then, assuming that efficiencies for the background and $\wp\wm$ signal are
sufficiently well understood that systematic uncertainties
effectively cancel in the ratio
of the above-threshold to the below-threshold rates, a very accurate $\mw$
determination becomes possible.

The dominant background derives from $e^+e^- \to
(Z/\gamma) (Z/\gamma)$ which is essentially
energy independent~\cite{lepii} below
180~GeV. For our present analysis we model the background as energy
independent, and accordingly assume that one measurement at an energy in the
range 140 to 150~GeV suffices to determine the background.

We analyze our ability to determine the $W$ mass via
just two measurements: one at center of mass energy $\sqrt{s}=161\gev$, just
above threshold, and one at $\sqrt{s}=150\gev$. 
The optimal $M_W$ measurement is obtained by expending about two-thirds
of the luminosity at $\sqrt{s}=161\gev$ and  one-third
at $\sqrt{s}=150\gev$.
Combining the three modes, an overall precision of
\begin{equation}
\Delta M_W = 6 \rm\ MeV
\end{equation}
should be achievable.

\begin{figure}[htb]
\leavevmode
\begin{center}
\epsfxsize=3.0in\hspace{0in}\epsffile{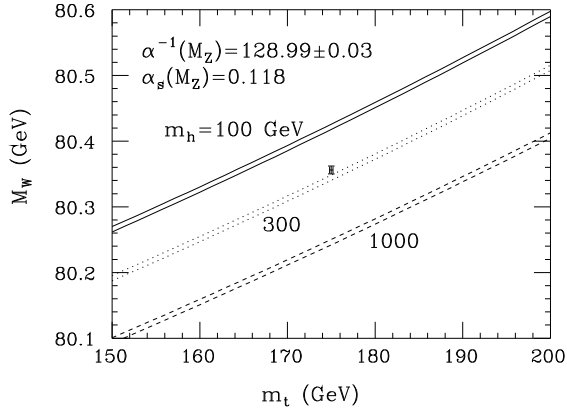}
\end{center}
\caption[]{\footnotesize\sf Correlation between $\protect M_W$ and $\protect m_t$
in the SM with QCD and electroweak corrections
for $\protect m_h=100, 300$ and 1000~GeV. The data point
and error bars illustrate the possible accuracy for
the indirect $m_h$ determination assuming
$M_W=80.356\pm 0.006$ GeV and $m_t=175\pm 0.2$ GeV.
The widths of the bands indicate the uncertainty
in $\protect \alpha(M_Z)$.}
\label{figure11}
\end{figure}

The combination of the measurements of
the masses $M_Z$, $M_W$ and $m_t$ to such high precision has dramatic
implications
for the indirect prediction of the mass of the Higgs boson and for other
sources of physics beyond the Standard Model. This is illustrated in
Fig.~\ref{figure11}. Assuming the current central values
of $M_W$, $m_t$,  $\alpha(M_Z)$ and $\alpha_s(M_Z)$, and that $L=10\fbi$
($100\fbi$) is devoted to the measurement of $\mt$ ($\mw$),
the mass of the SM Higgs
boson would be determined to be 260 GeV with an error of about $\pm 5$~GeV from
$\Delta m_t=200$~MeV at a fixed $M_W$,
and about $\pm 20$~GeV from $\Delta M_W=6$~MeV at a fixed $m_t$.
For $m_h=100$~GeV, the corresponding values would be
$\pm 2$~GeV  and $\pm 10$~GeV, respectively.
More generally, the $\Delta m_h$ value scales roughly like $m_h$.

%
\section*{Higgs Boson Measurement at the
$\mu^+\mu^-\to Z\lowercase{h}$ Threshold}

A very accurate determination of $\mh$ is obtained
by measuring the threshold cross section for the Bjorken Higgs-strahlung
process~\cite{bj} $\ell^+\ell^-\to Z\h$.
With integrated luminosity $L=100\fbi$, a
$1\sigma$ precision of order 45~MeV is possible for
a SM Higgs $\mh=100\gev$ at a $\mu^+\mu^-$ collider.
This error in $\mh$ is smaller than that achievable
via final state mass reconstruction for a typical detector, and would then
be the most accurate determination of $\mh$ at an $\epem$ collider.

The SM Higgs boson is easily discovered in the $Z\h$ production
mode by running the machine well above threshold, \eg\ at $\rts=500\gev$.
For $\mh\lsim 2\mw$ the dominant Higgs boson decay is to $b\anti b$
and most backgrounds can be eliminated by $b$-tagging.
The best means for measuring $\mh$
will be to first determine $\mh$ to within a few hundred MeV in
$\rts=500\gev$ running, which will also yield a precise
measurement of $\sigma(Z\h)$, and then reconfigure the collider
for maximal luminosity in the threshold energy region $\rts\approx \mz+\mh$.

\begin{figure}[htb]
\leavevmode
\begin{center}
\epsfxsize=3.0in\hspace{0in}\epsffile{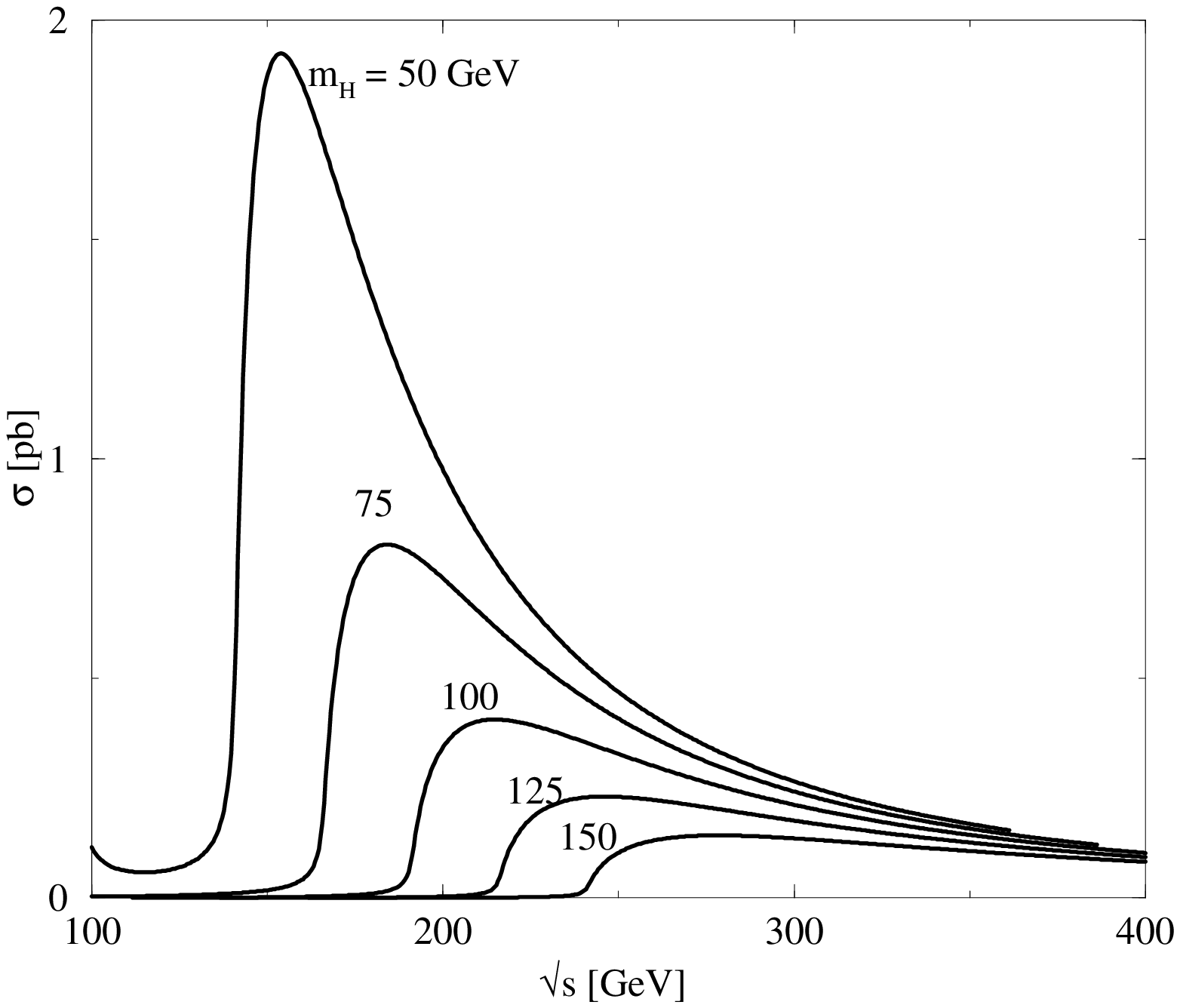}
\end{center}
\caption[]{\footnotesize\sf The cross section vs. $\protect\rts$
for the process
$\mu^+\mu^-\to Z^\star\h\to f\anti f\h$ for a range of Higgs masses.}
\label{zhcurve}
\end{figure}

In Fig.~\ref{zhcurve} we show the cross section for the Bjorken process
$\ell^+\ell^-\to Z\h$ for Higgs masses from 50 to 150~GeV. Since the threshold
behavior is $S$-wave, the rise in the cross section (which is a few tenths 
of a pb) in the threshold region
is rapid. 

In the ideal case that the normalization of the measured $Z\h$ cross section
as a function of $\rts$ can be precisely predicted, including
efficiencies and systematic effects,
sensitivity to the SM Higgs boson mass is maximized by
a single measurement of the cross section at $\rts=\mz+\mh+0.5$~GeV,
just above the real particle threshold. 
As an example of the precision that might be achieved, suppose $\mh=100$~GeV
and backgrounds are neglected.
The $Z\h$ cross section is 120~fb and is rising at a rate of 0.05~fb/MeV.
With $L=50\fbi$ and including an overall
($b$-tagging, geometric and event identification)
efficiency of $40\%$,
this yields $2.4\times 10^3$ events, or a measurement of the
cross section to about 2\%. From the slope of the cross section one concludes
that a $\mh$ measurement with accuracy of roughly 50~MeV is possible.

For a more precise estimate of the accuracy
with which $\mh$ can be measured, we employ $b$-tagging and cuts in order to
reduce the background to a very low level. These cuts and other systematic
uncertainties are discussed in more detail in Ref.~\cite{Zh}.
The background is very much smaller than the signal unless $\mh$
is close to $\mz$. We note that electroweak radiative corrections to the 
cross section are
estimated to be less than 1\% for $m_H\sim100$~GeV\cite{kniehl,HK}.
A precision of the SM Higgs mass determination to within 45\mev\
for $\mh=100\gev$ may be achievable at a muon collider.

\begin{figure}[htb]
\leavevmode
\begin{center}
\epsfxsize=5.0in\hspace{0in}\epsffile{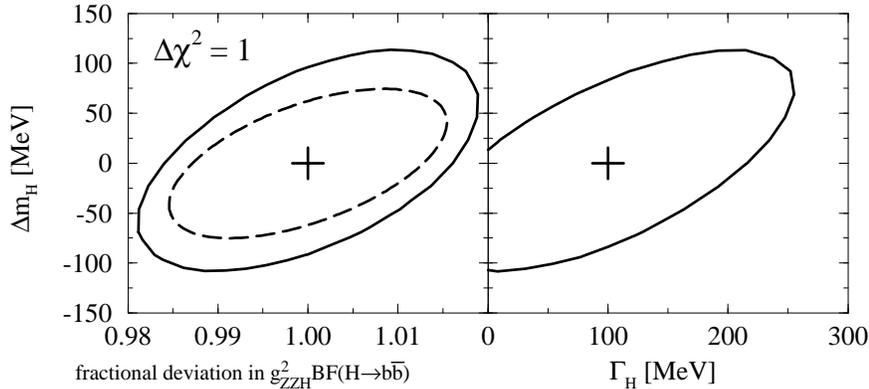}
\end{center}
\caption[]{\footnotesize\sf Solid curves show the $\Delta \chi ^2=1$ contours
for determining
the Higgs mass versus $g_{ZZ\h}^2\br(\h\to b\anti b)$, or versus
$\Gamma_H^{}$, by devoting
$100/3\fbi$ to each of the c.m.\ energies
$\protect\rts=\mz+\mh+0.5$~GeV, $\protect\rts=\mz+\mh+20$~GeV
and $\protect\rts=\mz+\mh-2$~GeV at a muon collider;
$b$-tagging and cuts 1)--4) are imposed
and initial state radiation and beam energy smearing are
included. A Higgs mass $m_H = 100$~GeV is assumed. The dashed curve shows the
$\Delta\chi^2=1$ contour that results when $\Gamma_H$ is negligibly small and
50~fb$^{-1}$ is devoted to each of the c.m.\ energies $\sqrt s = m_H + M_Z +
0.5$~GeV and $\sqrt s = m_H + M_Z  + 20$~GeV.}
\label{zhsin2}
\end{figure}

Outside the Standard Model the cross section generally depends on the Higgs
mass, the $ZZH$ coupling ($g_{ZZ\h}^{}$) and the total Higgs width 
($\Gamma_H$).
In order to simultaneously determine
$\mh$, $g_{ZZ\h}^{}$ and $\Gamma_H$,
measurements could be made at the three c.m.\ energies
$\rts=\mh+\mz+20\gev$,
$\rts=\mh+\mz+0.5\gev$, and
$\rts=\mh+\mz-2\gev$.
The solid curves in Fig.~\ref{zhsin2} show the statistical precision that can
be obtained for $\mh=100\gev$ in a three-parameter fit to
$\mh$, $g_{ZZ\h}^2\br(\h\to b\anti b)$ and $\Gamma_H$,
including smearing effects from bremsstrahlung, beamstrahlung
and beam energy spread at a muon collider,
using an integrated luminosity of 100/3~fb$^{-1}$ at each of the above three
values of $\rts$ in the threshold region.
The crosses in the center of the ellipses indicate the input values.
With a three-parameter fit,
the attainable error in $\mh$ is about $110\mev$ at the $1\sigma$ level.
The second panel in Fig.~\ref{zhsin2} shows that there is significant
sensitivity to the
Higgs width $\Gamma_H$ if it is of order 100 MeV.
If $\Gamma_H$ is very narrow ($\sim 3\mev$ is predicted in the SM)
then $\Delta\mh \sim\pm 80\mev$ is possible (see the dashed ellipse)
from a fit to
$\mh$ and $\sigma(Z\h)\br(\h\to b\anti b)$ by devoting 50~fb$^{-1}$ at each of
two c.m.\ energies, $\sqrt s = m_H + M_Z + 20$~GeV and $\sqrt s = m_H + M_Z +
0.5$~GeV.
Measurements that would simultaneously determine $\mh$,
$\sigma(Z\h)\br(\h\to b\anti b)$ and $\Gamma_H^{}$ could be done at
a level of accuracy that could distinguish a Standard Model
Higgs boson from its many possible (\eg\ supersymmetric) extensions.

\section*{Conclusion}

A muon collider offers an unparalleled opportunity for precision $W$ and
Higgs mass measurements in the respective threshold regions.
These measurements, however, require a collider that can deliver substantial
luminosity.

\section*{Acknowledgments}

I thank V.~Barger, J.~F.~Gunion and T.~Han for a pleasant collaboration
on the issues reported here.
This work was supported in part by the U.S. Department of Energy
under Grant
No.~DE-FG02-91ER40661.


\begin{references}

\bibitem{bbgh3} V.~Barger, M.S.~Berger, J.F.~Gunion and T.~Han,
Phys.\ Rev.\ {\bf D56}, 1714 (1997).

\bibitem{Zh} V. Barger, M. S. Berger, J. F. Gunion and T. Han,
Phys.\ Rev.\ Lett.\ {\bf 78}, 3991 (1997).

\bibitem{mnw} T.~Muta, R.~Najima and S.~Wakaizumi, Mod.\ Phys.\ Lett.\
{\bf A1}, 203 (1986).

\bibitem{coulomb} V.~S.~Fadin, V.~A.~Khoze and A.~D.~Martin, Phys.\ Lett.\
{\bf B311}, 311 (1993); V.~S.~Fadin, V.~A.~Khoze, A.~D.~Martin and
W.~J.~Stirling, Phys.\ Lett.\ {\bf B363}, 112 (1995); V.~S.~Fadin, V.~A.~Khoze,
A.~D.~Martin and A.~Chapovsky, Phys.\ Rev.\ {\bf D52}, 1377 (1995).

\bibitem{lepii} Z. Kunszt and W.J. Stirling {\it et al.},
hep-ph/9602352,
in {\it Proceedings of the Workshop on Physics at LEP2}, eds.\
G. Alterelli, T. Sjostrand and F. Zwirner, CERN Yellow Report
CERN-96-01 (1996), Vol.~1, p.~141; W.J. Stirling, Nucl. Phys.
{\bf B456}, 3 (1995).

\bibitem{bj} J.D.~Bjorken, {\it Proceedings of the Summer Institute on
Particle Physics}, ed. M.~Zipf (Stanford, 1976).

\bibitem{bbgh12} V.~Barger, M.S.~Berger, J.F.~Gunion and T.~Han,
Phys.\ Rev.\ Lett.\ {\bf 75}, 1462 (1995); 
Phys.\ Rept.\ {\bf 286}, 1 (1997).

\bibitem{bsss} E.~Boos, M.~Sachwitz, H.~J.~Schreiber and S.~Shichanin,
Int.\ J.\ Mod.\ Phys.\ {\bf A10}, 2067 (1995).

\bibitem{summary} {\it Higgs Boson Discovery and Properties},
J.F. Gunion \etal, {\it Proceedings of the 1996 Snowmass Workshop}.

\bibitem{kniehl} B.A. Kniehl, Z. Phys.\ {\bf C55}, 605 (1992).

\bibitem{HK} R. Hempfling and B. Kniehl, Z.~Phys.\ {\bf C59}, 263 (1993).

\end{references}
\end{document}